\documentclass[prl,superscriptaddress,twocolumn]{revtex4-2}
\usepackage{hyperref}
\usepackage{graphicx}
\usepackage{dcolumn}
\usepackage{bm}
\usepackage{amsfonts}
\usepackage{amsmath, amssymb}
\usepackage{float}
\usepackage[mathlines]{lineno}

\usepackage{color}
\usepackage[T1]{fontenc}
\newcommand{\neighbour}[1]{ \Gamma^{#1}}
\newcommand{\absDiff}[2]{|\mathbf{x}_k^{#1} - \mathbf{x}_k^{#2}|}
\newcommand{\unitVec}[2]{\frac{\mathbf{x}_k^{#1} - \mathbf{x}_k^{#2}}{|\mathbf{x}_k^{#1} - \mathbf{x}_k^{#2}|}}

\newcommand{\SFigStructure}{S2}

\newcommand{\particlewise}{individual-particle }
\newcommand{\Particlewise}{Individual-particle }
\newcommand{\pairwise}{pairwise }
\newcommand{\Pairwise}{Pairwise }

\title{}

\newcommand{\nyuphysics}{Center for Soft Matter Research, Department of Physics, New York University, New York 10003, USA}

\newcommand{\nyusimons}{Simons Center for Computational Physical Chemistry, Department of Chemistry, New York University, New York 10003, USA}
\newcommand{\nyucourant}{Courant Institute of Mathematical Sciences, New York University, New York 10003, USA}
\newcommand{\nyucns}{Center for Neural Science, New York University, New York 10003, USA}

\begin{document}

\title{Absorbing state dynamics of stochastic gradient descent}

\author{Guanming Zhang}
\affiliation{\nyuphysics}
\affiliation{\nyusimons}
\email{gz2241@nyu.edu}
\author{Stefano Martiniani}
\affiliation{\nyuphysics}
\affiliation{\nyusimons}
\affiliation{\nyucourant}
\affiliation{\nyucns}
\email{sm7683@nyu.edu}
\begin{abstract}
Stochastic gradient descent (SGD) is a fundamental tool for training deep neural networks across a variety of tasks. In self-supervised learning, different input categories map to distinct manifolds in the embedded neural state space. Accurate classification is achieved by separating these manifolds during learning, akin to a packing problem. We investigate the dynamics of ``neural manifold packing'' by employing a minimal model in which SGD is applied to spherical particles in physical space. In this model, SGD minimizes the system's energy (classification loss) by stochastically reducing overlaps between particles (manifolds). We observe that this process undergoes an absorbing phase transition, prompting us to use the framework of biased random organization (BRO), a nonequilibrium absorbing state model, to describe SGD behavior. We show that BRO dynamics can be approximated by those of particles with linear repulsive interactions under multiplicative anisotropic noise. Thus, for a linear repulsive potential and small kick sizes (learning rates), we find that BRO and SGD become equivalent, converging to the same critical packing fraction $\phi_c \approx 0.64$, despite the fundamentally different origins of their noise. This equivalence is further supported by the observation that, like BRO, near the critical point, SGD exhibits behavior consistent with the Manna universality class. Above the critical point, SGD exhibits a bias towards flatter minima of the energy landscape, reinforcing the analogy with loss minimization in neural networks. 
\end{abstract}

\maketitle


Stochastic gradient descent (SGD) is widely used in machine learning and optimization, particularly for training deep neural networks (DNN), due to its efficiency and ability to handle large datasets via batch optimization \cite{bubeck2015convex, bishop2006pattern,bishop2024deep,bengio2012practical,rumelhart1986learning,bottou2003large,bengio2012deep}. When training a DNN, the discrepancy between the data and the model's prediction is given by a loss function \cite{wang2020comprehensive,janocha2017loss,geiger2019jamming,pellegrini2020analytic,krizhevsky2012imagenet,simonyan2014very}. SGD iteratively updates the model's weights by performing gradient descent on randomly selected batches of data to minimize the loss. Despite being overparameterized, DNNs trained with SGD exhibit surprisingly strong generalization performance. While a comprehensive theory of deep learning is still under development \cite{jacot2018neural,gerbelot2024rigorous,li2017stochastic,li2019stochastic,rotskoff2022trainability,yang2023stochastic,mei2018mean}, it is hypothesized that this generalization ability arises from the interplay between training dynamics and the geometric structure of the loss landscape of DNNs \cite{azizian2024long,li2018visualizing,ainsworth2022git}. Therefore, developing a deeper understanding of SGD dynamics is of critical importance.

\begin{figure}[b]
    \includegraphics[width=0.3\textwidth]{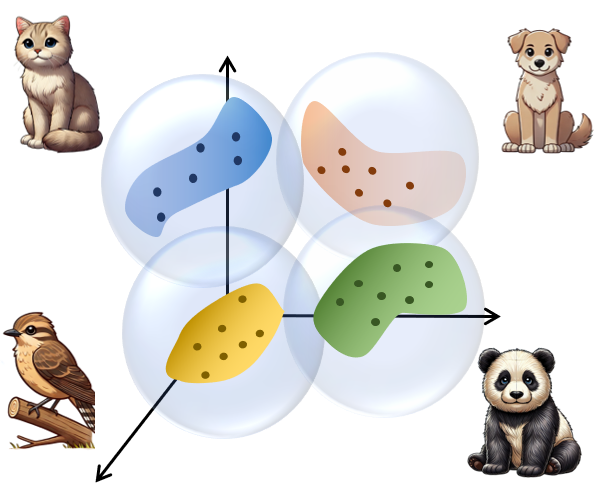}
    \caption{Neural manifold packing. Each manifold in the embedded neural state space represents an animal class. When the manifolds are enclosed by spheres, the classification problem reduces to a sphere packing problem.}
    \label{fig:schematics-representaion}
\end{figure}

In the neural state space, points representing neural responses to inputs from the same class collectively form a distinct manifold. Neural activity measurements of large populations of neurons in the mouse visual cortex have shown that neural manifolds are both correlated and high-dimensional \cite{stringer2019high, stringer2019spontaneous, kafashan2021scaling} (Fig.~\ref{fig:schematics-representaion}). Classification aims to distinguish these manifolds using a readout network (or decoder). If the finite-sized neural manifolds representing “cats” and “dogs” were to overlap, some dogs would be perceived as cats and vice versa, resulting in decoding errors. Consequently, classification can be seen as a packing problem in neural state space. A simplistic solution to this problem would be to shrink each manifold to a single point, eliminating overlaps entirely. However, if the “cats” manifold collapsed to a point, all cats would be perceived as identical, making the representation inadequate for other downstream tasks requiring detailed differentiation.

An example of how deep learning techniques explicitly manipulate the neural state space is self-supervised learning (SSL). SSL enables models to learn meaningful representations without large labeled datasets by clustering the embedded representations of an image and its augmentations (i.e., distortions) while simultaneously separating the representations of different images. Recent advances show that SSL can match or even outperform supervised learning across various tasks \cite{chen2020simple,he2020momentum,goyal2022vision}.

In a simplified theoretical treatment, where neural manifolds are conceptually enclosed within spheres, the training problem reduces to finding configurations where these spheres do not overlap (Fig.~\ref{fig:schematics-representaion}). In the limit of a large number of classes encoded by the responses of a downstream layer of 3 neurons (i.e., in $3d$ embedded neural state space), this problem reduces to the classical sphere packing problem studied in physics \cite{Bernal1960packing,behringer2018physics}, and, in higher dimensions, to sphere packing problems considered in mathematics and information theory \cite{conway2013sphere}. 

Manifold capacity theory (MCT) quantifies the capacity of a linear decoder for \textit{binary} classification of neural manifolds whose dimensionality is much lower than that of the embedding space \cite{chung2018classification,chung2016linear}, with recent work extending it to binary classification with nonlinear decoders \cite{mignacco2024nonlinear}. Unlike MCT, we focus on the learning dynamics of the encoder for the general $n$-ary classification problem (i.e., $n \gg 2$), which is the relevant setting for SSL, rather than on the binary capacity of the decoder. Furthermore, we consider manifolds that span the embedding space, consistently with recent experimental evidence on the structure of neural representations in the visual cortex \cite{stringer2019high, stringer2019spontaneous, kafashan2021scaling}. Our analysis is also different from previous studies that proposed a more tentative link between classification and packings in the weight space \cite{geiger2019jamming}. Instead, we argue that the connection to neural state space, inspired by contemporary systems neuroscience, is more direct and concrete.

A particularly insightful perspective on the packing problem is offered by ``random organization'' models. Originally introduced to model the dynamics of driven colloidal suspensions, \cite{corte2008random,wilken2020hyper,tjhung2015hyperuniform} these ``absorbing state models'' exhibit a nonequilibrium phase transition from an absorbing (inactive) state, where all geometric constraints are satisfied (i.e., no particle overlaps), to an active steady state with unresolved constraints. The dynamics of random organization models are prescribed by local stochastic rules. For instance, in Biased Random Organization (BRO), the system evolves by randomly displacing overlapping particles away from one another. Interestingly, BRO displays several characteristics of random close packing (RCP), such as a critical point at volume fraction $\phi_c \approx 0.64$, while also displaying Manna universality scaling \cite{wilken2021random,wilken2023dynamical}.  When particle updates in BRO involve nonreciprocal, collective kicks from all neighbors that do not strictly conserve center of mass, we refer to this variant as ``\particlewise BRO''. Conversely, if the kicks are reciprocal and the updates conserve center of mass,  we denote it as ``\pairwise BRO''. 
Recent studies have demonstrated that \pairwise BRO exhibits long-range order in $2d$, violating the Mermin-Wagner theorem \cite{galliano2023two}, and that density fluctuations in the active phase are anomalously suppressed \cite{hexner2017noise}.

BRO's random kicks are characterized by an intrinsic bias: particles repel each other even without an explicitly defined potential. This repulsion is induced by a multiplicative noise process that triggers kicks only when particles overlap. Interestingly, these dynamics share similarities with those of stochastic gradient descent (SGD) \cite{mignacco2020dynamical,mignacco2022effective,rotskoff2022trainability}. However, a key distinction between BRO and SGD lies in the origin of their noise: in BRO, noise stems from variations in kick sizes, whereas in SGD, it arises from the randomness of batch selection. Whether these processes can be unified within a common framework remains an open question -- one we address in this work, demonstrating that such unification is indeed possible.

In this work, we examine the absorbing state dynamics of neural manifold packing through BRO, SGD, and a related algorithm known as ``random coordinate descent'' (RCD), specifically in the limit of a large number of classes (particles) within a three-dimensional embedded neural state space. This approach allows us to establish direct connections between our findings and the physics literature on BRO. We begin by showing that the discrete \particlewise and \pairwise BRO dynamics can be effectively approximated by the dynamics of linearly repulsive particles driven by anisotropic multiplicative noise, defined at the single-particle and particle-pair levels, respectively. We then demonstrate both analytically and numerically that, for a system of linear-repulsive particles, RCD -- a gradient descent method that updates randomly selected particle coordinates -- and energy-based SGD, where the energy function is stochastically constructed via random pairwise selection, closely approximate \particlewise and \pairwise BRO, respectively.
Across all schemes, as the kick size (or, equivalently, the learning rate for RCD and SGD) approaches zero in three dimensions, all methods converge to the same critical packing fraction, $\phi_c \approx 0.64$, which coincides with RCP \cite{kamien2007random,wilken2020hyper,donev2005unexpected,desmond2009random, anzivino2023estimating}. In all cases, the critical behavior near the critical point is in the Manna universality class. This consistency holds regardless of batch size, even in the zero noise limit of RCD and SGD (i.e., gradient descent). These results suggest that the distinctive noise characteristics of the different processes are unimportant near criticality. In contrast, in the active phase at higher packing fractions, SGD with smaller batch sizes favors flat minima, whereas RCD is biased towards sharp minima.

\begin{figure}[b]
    \includegraphics[width=0.48\textwidth]{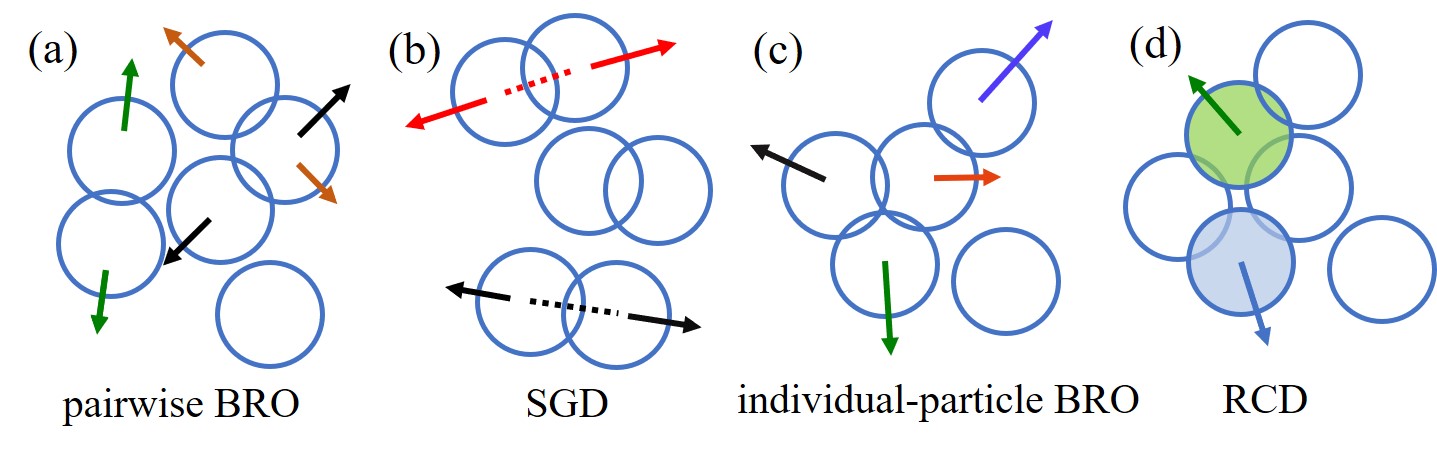}
    \caption{ Schematics of 
    (a) \Pairwise BRO: each active pair is kicked randomly by opposite equally-sized vectors.
    (b) SGD: two of three active pairs (dashed lines) are selected to form a batch, and gradient descent is performed individually for each pair.
    (c) \Particlewise BRO: each active particle is randomly kicked away collectively from all overlapping particles. 
    (d) RCD: the coordinates of two particles (shaded) out of five are selected to form a batch, and gradient descent is performed for each of the batch particles with respect to all interactions.}
    \label{fig:schematics}
\end{figure}

\paragraph{\bf{\Pairwise BRO as linear-repulsive particles driven by anisotropic multiplicative noise.}}
In the \pairwise BRO model, each pair of overlapping particles is displaced by equal-magnitude vectors in opposite directions along the line connecting their centers. The magnitude of these displacements is sampled uniformly from the range $[0, \epsilon]$, where $\epsilon$ denotes the kick size. In each iteration, overlapping particles, which are said to be ``active'', are updated, while isolated particles, termed ``inactive'', remain unchanged.  This process is repeated until the system either reaches an absorbing state with no active particles or attains an active steady state where the fraction of active particles stabilizes around some value $0<f_a \leq 1$. Fig.~\ref{fig:schematics}(a) illustrates the dynamics of \pairwise BRO (See Supplementary Material for the algorithmic description). The update rule for active particle $i$ at iteration $k$ then reads \cite{wilken2021random,ness2020absorbing},
\begin{equation}
        \mathbf{x}_{k+1}^{i} =\mathbf{x}_{k}^{i} + \epsilon  \sum_{j \in \Gamma_k^i} u^{ji}_k \Delta \mathbf{x}_k^{ji},
     \label{eq:pairwise-bro-step}
\end{equation}
where $\mathbf{x}_k^{i}$ is the position of particle $i$ at time step $k$, and $\Delta \mathbf{x}_k^{ji} = -  (\mathbf{x}_k^j - \mathbf{x}_k^i)/|(\mathbf{x}_k^j - \mathbf{x}_k^i)|$ is the direction of the kick from particle $j$ to $i$. $\epsilon$ is the kick size, $u_k^{ji}$ is a ``reciprocal'' noise sample drawn uniformly from $[0,1]$ such that $u_k^{ji} = u_k^{ij}$. $\Gamma_k^i = \{j |\  |\mathbf{x}_k^j - \mathbf{x}_k^i| < 2R \}$ is the set of all particles overlapping with particle $i$ at the iteration $k$, and $R$ is the particles' radius.

To derive a stochastic approximation of these dynamics \cite{li2019stochastic,hu2019diffusion}, we decompose the biased random kick on each pair into its mean value, $\mathbf{g}_k^{ji} = \mathbb{E}[ \Delta \mathbf{x}_k^{ji} u_k^{ji} ] = \Delta \mathbf{x}_k^{ji}/2$, and the fluctuation around the mean, $\mathbf{z}_k^{ji} = u_k^{ji} \Delta \mathbf{x}_k^{ji} - \mathbf{g}_k^{ji}$. We then approximate the random fluctuations by a Gaussian random variable, $\boldsymbol{\eta}_k^{ji} \sim \mathcal{N}(\mathbf{0}, \boldsymbol{\Sigma}^{ji})$, with covariance $\mathbf{\Sigma}^{ji} = \mathbb{E}[\mathbf{z}_k^{ji} \mathbf{z}_k^{ji^T}] = \frac{1}{3}\mathbf{g}_k^{ji} \mathbf{g}_k^{ji^T}$. As a result, Eq.~\ref{eq:pairwise-bro-step} is approximated by the process
\begin{equation}
    \begin{aligned}
        \mathbf{x}_{k+1}^i =\mathbf{x}_{k}^i + \epsilon \sum_{j \in \neighbour{i}_k}(\mathbf{g}_k^{ji} + \boldsymbol{\eta}_k^{ji}), 
        \label{eq:bro_approx1}
    \end{aligned}
\end{equation}
We note that the $\mathbf{g}_k^{ji}$ is equivalent to the negative gradient of the pairwise potential energy of short-ranged, linear-repulsive particles, 
\begin{equation}
    \begin{aligned}
       \mathbf{g}^{ji}_k &= -\nabla^i U(|\mathbf{x}_k^i-\mathbf{x}_k^j|)\\
        U(\mathbf{r}) &= 
    \begin{cases}
      \frac{1}{2}(2R - |\mathbf{r}|), & \text{if}\ |\mathbf{r}| < 2R \\
      0, & \text{otherwise}
    \end{cases} 
    \end{aligned}
    \label{eq:linear_pot}
\end{equation}
Thus, we can define the total potential energy $V = \sum_i \sum_{j>i} U(|\mathbf{x}_k^i - \mathbf{x}_k^j|)$, such that
$\sum_{j} \mathbf{g}^{ji}_k = - \nabla^i V$, and rewrite Eq.~\ref{eq:bro_approx1} to arrive at the stochastic approximation, 
\begin{equation}
   \begin{aligned}
       \mathbf{x}_{k+1}^{i} &=  \mathbf{x}_{k}^{i} - \epsilon \nabla^{i} V + \frac{\epsilon}{\sqrt{3}} \sum_{j} \sqrt{\boldsymbol{\Lambda}^{ji}} \cdot \boldsymbol{\xi}_k^{ji},
    \text{where } \\
    \boldsymbol{\Lambda}^{ji} &=  \nabla^{i} U(|\mathbf{x}_k^i-\mathbf{x}_k^j|) \ \nabla^{i}  U(|\mathbf{x}_k^i-\mathbf{x}_k^j|)^T
   \end{aligned}
   \label{eq:pairwise-bro-sto-approx}
\end{equation}
 $\boldsymbol{\xi}_k^{ji}$ is standard Gaussian noise modelling the reciprocal, pairwise kick from $j$ to $i$, such that  $\boldsymbol{\xi}_k^{ji} = -\boldsymbol{\xi}_k^{ij}$. $\sqrt{\boldsymbol{\Lambda}^{ji}/3}$ is the matrix square root of the covariance matrix $\mathbf{\Sigma}^{ji}$ , and serves to scale and project the noise along the lines connecting the particles' centers. Eq.~\ref{eq:pairwise-bro-sto-approx} demonstrates that the BRO dynamics are equivalent to those of linearly repulsive particles driven by anisotropic multiplicative noise. This marks the first key result of our study. 
 
 In the limit of small kick size, the BRO dynamics can be further approximated as a continuous-time process described by the stochastic differential equation \cite{stephan2017stochastic,li2017stochastic,li2019stochastic}, 
 \begin{equation}
      \mathrm{d}\mathbf{x}^{i}(t) = - \frac{\epsilon}{\tau} \nabla^{i} V \ \mathrm{d}t+ \frac{\epsilon}{\sqrt{3\tau}} \sum_{j} \sqrt{\boldsymbol{\Lambda}^{ji}(t)} \cdot \mathrm{d}\boldsymbol{W}^{ji}(t)
      \label{eq:pairwise-bro-sme}
 \end{equation}
 where $\mathbf{x}^{i}(t)$ and $\boldsymbol{\Lambda}^{ji}(t)$ are the continuous-time counterparts of the discrete-time variables in Eq.~\ref{eq:pairwise-bro-sto-approx} and $\tau$ is the time scale measuring the time elapsed between two discrete steps (see Supplemental Materials).

\paragraph{\bf{SGD approximates \pairwise BRO dynamics.}}
 In the context of interacting particle systems, we introduce \textit{energy-based SGD}, inspired by the SGD method widely used in machine learning. During each iteration $k$, an active pair (a pair of particles $i$ and $j$ such that $|\mathbf{x}_k^i - \mathbf{x}_k^j| \leq 2R$) is selected with probability $b_f$, equal to the fraction of pairs in a batch. Each particle in the selected pair is then moved by equal but opposite displacements $-\alpha \nabla^i U(|\mathbf{x}_k^i-\mathbf{x}_k^j|)$ and $-\alpha \nabla^j U(|\mathbf{x}_k^i-\mathbf{x}_k^j|)$, where $\alpha$ is the learning rate. We choose the interaction potential $V = \sum_i \sum_{j>i} U(|\mathbf{x}_k^i - \mathbf{x}_k^j|)$ to be pairwise, short-ranged and linearly-repulsive, as in Eq.~\ref{eq:linear_pot}, to be consistent with the \pairwise BRO process. However, the potential energy $V$ minimized by SGD need not be the same as in Eq.~\ref{eq:linear_pot}, in fact other choices of pairwise potentials (e.g., Lennard-Jones \cite{lennard1924determination}, WCA \cite{weeks1971role}, etc.) can be used to construct a valid SGD process. 
 
 Selecting a batch of active pairs corresponds to selecting a subset of terms in the sum defining $V$, which is a sum over pairwise interactions. This is analogous to selecting a subset of terms from the contrastive loss function $\mathcal{L}(\mathcal{D}) = \sum_{\mathbf{x}_i,\mathbf{x}_j \in \mathcal{D}} L[\hat{f}_{\boldsymbol{\theta}}(\mathbf{x}_i), \hat{f}_{\boldsymbol{\theta}}(\mathbf{x}_j)]$ where $\mathcal{D} = \{ \mathbf{x}_1,  \dots, \mathbf{x}_N \}$ is a dataset, $\hat{f}_{\boldsymbol{\theta}}(\mathbf{x}_j)$ represents the model output, parametrized by $\boldsymbol{\theta}$. Selecting a pair of particles is similar to selecting a pair of data samples. The contrastive loss 
$\mathcal{L}$ maximizes the separation between the model’s responses to different samples via adapting $\boldsymbol{\theta}$, thereby reducing manifold overlap, mirroring the role of repulsive interactions in the stochastic particle dynamics. 
 
 To approximate the stochastic updates in SGD, we determine the probability distribution for selecting a particle pair and then calculate the mean and covariance of the descent step. Following the same stochastic approximation approach used in Eq.~\ref{eq:bro_approx1}, we arrive at the following stochastic approximation of the SGD dynamics (see Appendix for details)
 \begin{equation}
    \mathbf{x}_{k+1}^i = \mathbf{x}_k^i - \alpha b_f \nabla^i V
     + \alpha \sqrt{b_f - b_f^2} \sum_{j \in \neighbour{i}_k} \sqrt{\boldsymbol{\Lambda}^{ji}} \cdot \boldsymbol{\xi}_k^{ji},
     \label{eq:sgd-sto-approx} 
\end{equation}
where $\boldsymbol{\xi}_k^{ji}$ is the standard Gaussian noise satisfying reciprocity, $\boldsymbol{\xi}_k^{ji} = -\boldsymbol{\xi}_k^{ij}$. The noise term in Eq.~\ref{eq:sgd-sto-approx} has the same functional form as Eq.~\ref{eq:pairwise-bro-sto-approx}, illustrating the shared stochastic nature for SGD and BRO. If we set $\alpha = \epsilon/b_f$ and $b_f = 3/4$, where $\epsilon$ is the kick size in the BRO dynamics, then Eq.~\ref{eq:sgd-sto-approx} matches Eq.~\ref{eq:pairwise-bro-sto-approx} exactly. This marks the second significant result of our study. 

Again, in the limit of infinitesimal learning rate, the SGD dynamics approximates the continuous time stochastic differential equation,
\begin{equation}
    \mathrm{d} \mathbf{x}^i(t) = - \frac{\alpha b_f}{\tau} \nabla^i V \ \mathrm{d}t
     +  \frac{\alpha}{\sqrt{\tau}}\sqrt{b_f - b_f^2} \sum_{j} \sqrt{\boldsymbol{\Lambda}^{ji}(t)} \cdot \mathrm{d} \boldsymbol{W}^{ji}(t).
     \label{eq:sgd-sme} 
\end{equation}

\paragraph{\bf{RCD approximates \particlewise BRO dynamics.}}
An alternative implementation of BRO applies particle-wise kicks at each iteration. In this approach, each overlapping neighbor of particle $i$ contributes a unit kick, and the resultant vector is scaled by a random number drawn from a uniform distribution over $[0,\epsilon]$, which then displaces particle $i$. We find that such a process approximates random coordinate descent (RCD) \cite{wright2015coordinate}, where each active particle has probability $b_f$ to be selected and moved along its gradient by $-\alpha \nabla^i V$ (see Appendix and Supplementary Material). Specifically, when $\alpha=\epsilon/b_f$ and $b_f=3/4$, the stochastic approximations for \particlewise BRO and RCD are equivalent.

\paragraph{\bf{Critical behavior.}}
We numerically compare the dynamics of BRO, RCD, SGD and their stochastic approximations, finding that in the limit of small kick sizes, their mean square displacements agree, and their critical packing fractions converge, $\phi_c \xrightarrow{} \phi_{RCP} \approx 0.64$ as $\epsilon \xrightarrow{} 0$ (or learning rate $\alpha  \xrightarrow{} 0$ for RCD and SGD), see Appendix and Supplementary Material for details. 

Thus, it becomes compelling to investigate whether SGD and RCD, like BRO, belong to the Manna (conserved direct percolation) universality class. We measure the steady state ``activity'', $f^a_\infty$, and the relaxation time $\tau_r$. The activity $f^a_k$ is defined as the ratio between the number of overlapping particles and the total number of particles at time step $k$, and $\tau_r$ measures the time for the system to reach the steady state. On the absorbing side, $\phi < \phi_c$, $\tau_r$ is the time required for the system to absorb, $f^a_{\tau_r} = 0$. On the active side, $\phi > \phi_c$, where the system never reaches an absorbing state and evolves forever, we estimate $\tau_r$ as the first time that the activity drops to within $(f_k^a - f^a_{\infty})/f^a_{\infty} < 0.1$ \cite{galliano2023two}. Random organization models display critical scalings $f^a_{ \infty} \sim (\phi - \phi_c)^{\beta}$ and  $\tau_r \sim |\phi - \phi_c|^{-\nu_{\parallel}}$ in the proximity of $\phi_c$, where $\beta = 0.84$ and $\nu_{\parallel} = 1.08$ are the Manna exponents \cite{martiniani2019quantifying, wilken2021random,tjhung2016criticality,wilken2020hyper}.
\begin{figure}[t]
    \centering
    \includegraphics[width=0.48\textwidth]{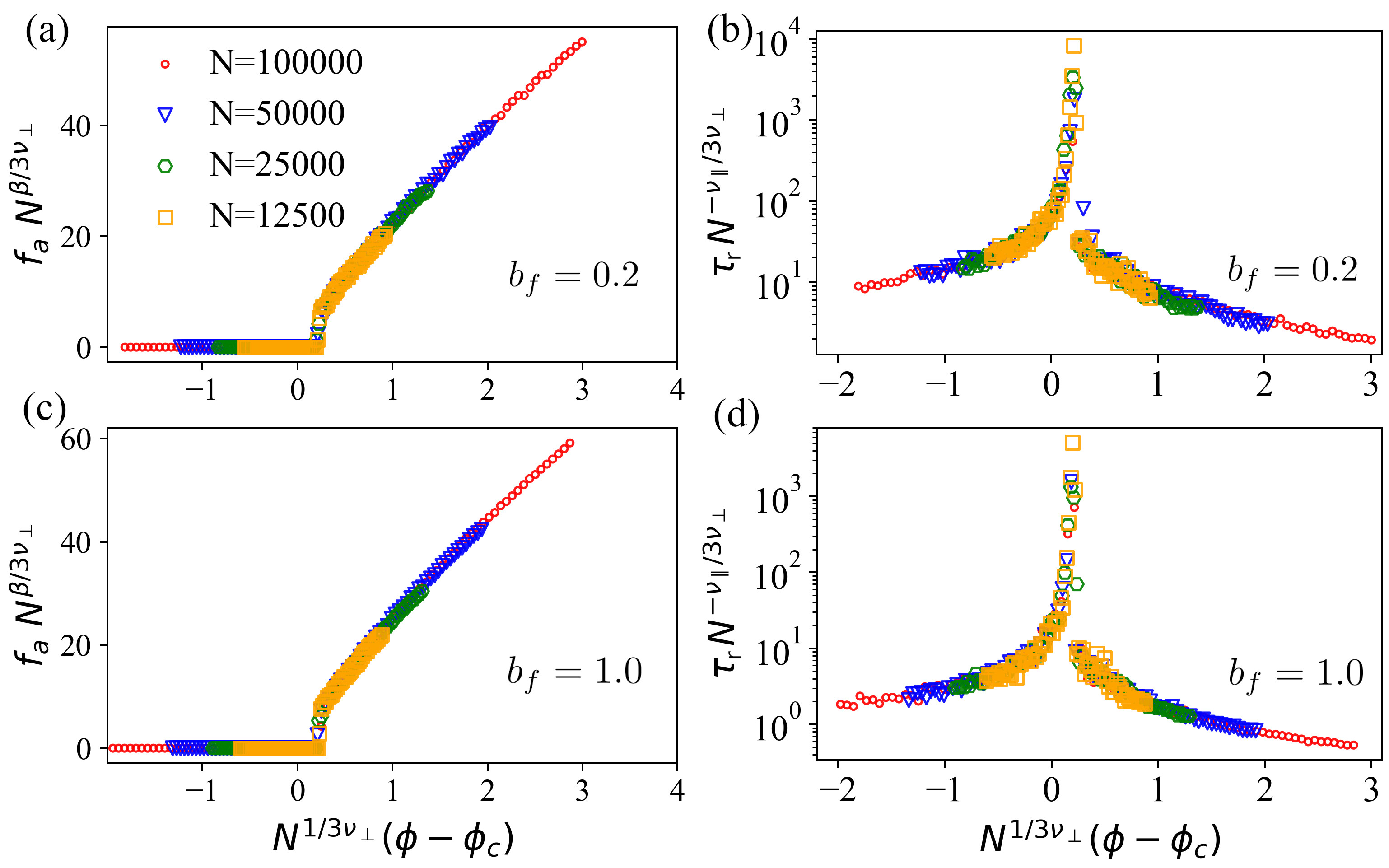}
    \caption{Finite size scaling analysis for SGD: Steady-state activity as a function of packing fraction, $\phi$, for (a) $b_f=0.2$ and (c) $b_f=1.0$. Relaxation time as a function of packing fraction for (b) $b_f=0.2$ and (d) $b_f=1.0$. $\nu_{\parallel} = 1.08, \beta = 0.84$ and $\nu_{\bot} = 0.59$ are Manna exponents in $3d$. 
    }
    \label{fig:exponents}
\end{figure}

We test this hypothesis by performing finite size scaling analysis for the activity and relaxation time of RCD and SGD at different batch fractions, $b_f$, assuming the Manna exponents \cite{Henkel2009Non} and fitting $\phi_c$ using Pyfssa \cite{sorge2015pyfssa}, see Fig.~\ref{fig:exponents} (and Supplementary Material). 
The data collapses are convincing both for RCD and SGD, irrespective of batch fractions ($b_f$), which corresponds to different noise levels. When $b_f = 1.0$, both RCD and SGD reduce to gradient descent (GD), corresponding to the zero noise limit. Consequently, the critical behavior for RCD, SGD and GD are consistent with the Manna universality class. However, it is important to note that GD is not an absorbing state model, as it ``absorbs'' (i.e., reaches a static steady state) on both sides of $\phi_c$. For $\phi > \phi_c$, GD settles into an overlapping state (a minimum of the potential $V$) rather than reaching an evolving steady state like SGD. To sample multiple ``active states'' using GD one needs to generate multiple relaxations from random initial conditions.
 
\begin{figure}[t]
    \includegraphics[width=0.48\textwidth]{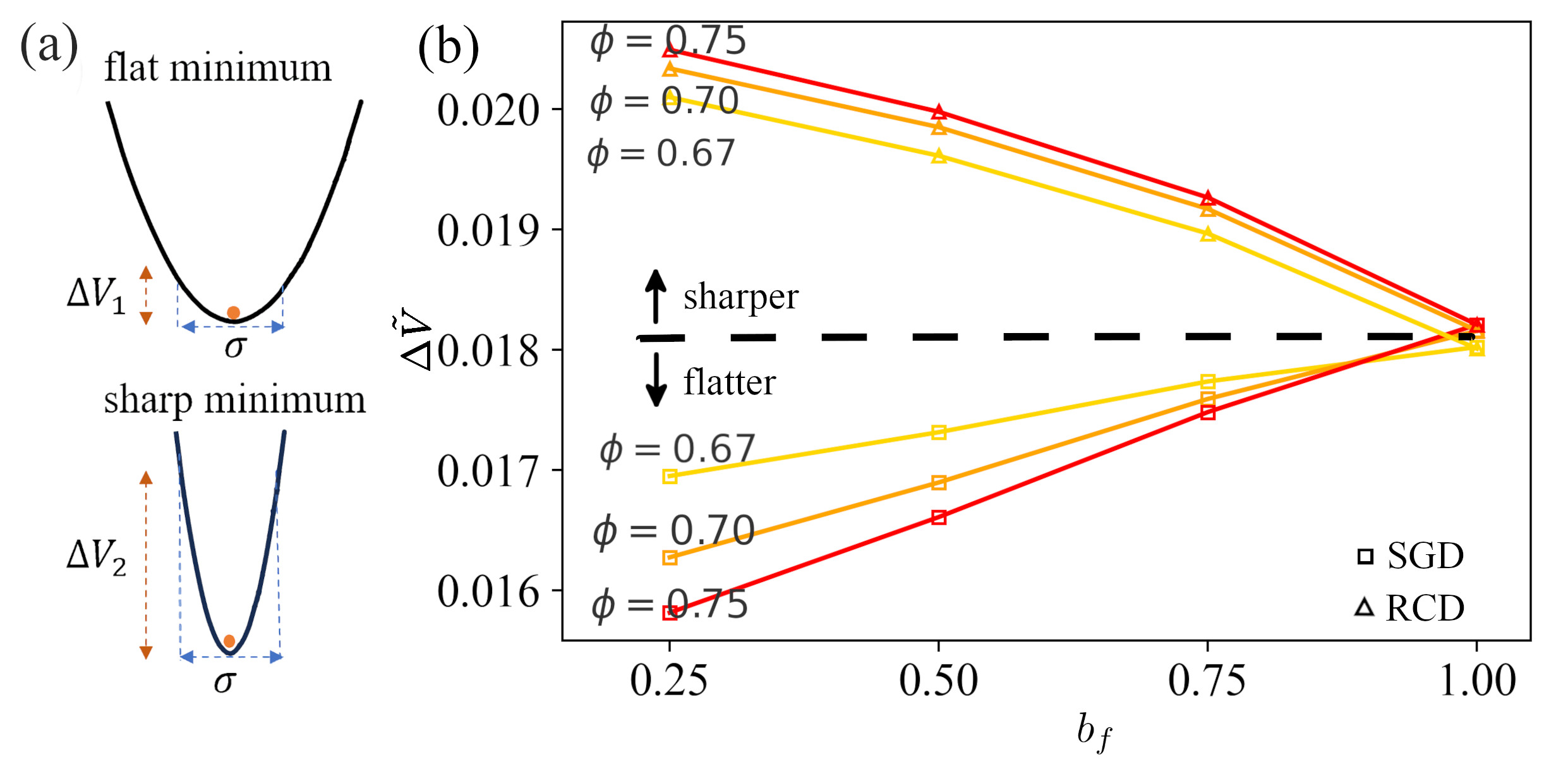}
    \caption{ Energy fluctuations of SGD steady state configurations under positional perturbations. (a) When the perturbation is of order $\sigma$, the energy fluctuation for the flat minimum ($\Delta V_1$) is smaller than for the sharp minimum ($\Delta V_2$). (b) Dimensionless energy fluctuation $\Delta \Tilde{V}$ as a function of batch fraction, $b_f$, for a number of packing fractions, $\phi$. For $b_f=1.0$, SGD reduces to gradient descent (GD). Each curve is averaged over $8$ independent simulations, and each steady state configuration is perturbed 4000 times to measure $\Delta \Tilde{V}$ with $\Tilde{\sigma}=0.03$.}
    \label{fig:flatness}
\end{figure}

\paragraph{\bf{Flatness of minima in the active phase of RCD and SGD.}} We have shown that RCD and SGD behave similarly near the critical point (Figs.~{\ref{fig:exponents}, \ref{fig:approx_phic}(b)} and Supplementary Material).
Here, we explore their behavior above the critical point, examining the steady state energy susceptibility of the two processes.  We perturb the steady state particle positions $\mathbf{X}_{T} = [\mathbf{x}_T^1,\mathbf{x}_T^2 , ... \mathbf{x}_T^N] $, at the last time step $k=T$, by the Gaussian noise, $\delta \mathbf{X} \sim \mathcal{N}(\mathbf{0},\sigma^2\mathbf{I})$, and calculate the average energy fluctuations $\Delta V = \langle V(\mathbf{X}_{T} + \delta \mathbf{X}) - V(\mathbf{X}_{T}) \rangle_{\delta \mathbf{X}}$ . Since $\mathbf{X}_{T}$  is near a local minimum, higher $\Delta V$ indicates greater susceptibility to positional perturbations. $\Delta V$ serves as a measure of the flatness of the minima, as illustrated in Fig.~\ref{fig:flatness}(a). If $V$ is second-order smooth, $\Delta V$ scales as the trace of the Hessian, $\Delta V \propto Tr(H)$ (see Supplemental Material) and it has been used to quantify the flatness of the loss landscape of DNNs \cite{jastrzkebski2017three}.

We compare the SGD and RCD schemes by perturbing the steady-state configuration with the same dimensionless noise variance, $\Tilde{\sigma} = \sigma / R$, and calculate the dimensionless energy fluctuation $\Delta \Tilde{V} = \Delta V/ U_0$ where $U_0 \equiv U(r=0) $ measures the energy scale for the pairwise repulsion. Fig.~\ref{fig:flatness}(b) demonstrates that for RCD, smaller batch sizes lead to sharper minima. Conversely, SGD is biased towards flatter minima for smaller batch sizes. This qualitative difference stems from their distinct stochastic dynamics. In RCD, each particle's displacement follows its local gradient to minimize the total potential energy $V$. In contrast, at each step, SGD selects a subset of active pairs to define a random partial energy landscape, $V'$, with respect to which it minimizes. This process, similar to SGD in neural networks, favors flat minima \cite{jastrzkebski2017three,xie2020diffusion,keskar2016large,huang2020understanding}. Our findings suggest that using small batch sizes for the random construction of the partial energy landscape, as in SGD, is essential for finding flat minima. This cannot be achieved by following the gradient of the total energy, as done in RCD.

\paragraph{\bf{Discussion.}} Our work demonstrates that displacement noise in BRO and selection noise in SGD play similar roles. Despite their different sources of randomness, all these processes converge to the same limiting packing fraction $\phi_c \rightarrow 0.64$ as the kick size, $\epsilon \xrightarrow[]{} 0 $ (or learning rate $\alpha \xrightarrow[]{} 0$ for SGD/RCD). Note that $\phi_c \approx 0.64$ represents the upper limit on the maximum capacity achievable by SGD. Although denser packing exists -- such as face-centered cubic packing ($\phi_{FCC} = \pi/\sqrt{18} \approx 0.74$) \cite{hales1998overview, zaccone2022explicit}) -- these configurations are not found by SGD when starting from random initial conditions. Greater maximum capacity, could be achieved by tuning the initial conditions or by applying SGD to systems of polydisperse spheres \cite{anzivino2023estimating} or ellipsoids \cite{donev2004improving,rocks2023structure}, which more closely approximate neural manifolds. Near the critical point, the details of different noise protocols become negligible, with all schemes exhibiting behavior consistent with the Manna universality class (a property that we expect to hold for polydisperse spheres and ellipsoids \cite{shen2025personal}). However, above the critical packing fraction, $\phi > \phi_c$, the nature of the random process used to construct the gradients significantly affects the type of minima reached by the algorithm. Notably, only SGD exhibits a bias towards flat minima, reinforcing our analogy with loss minimization in neural networks. 

This work paves the way for understanding how neural manifolds could evolve through a mechanism equivalent to SGD. Our results can be extended to higher dimensions, corresponding to classification problems in neural state space. We hypothesize that, as the number of classes (equivalent to the number of particles in our model) approaches infinity and the number of dimensions $d\geq 4$, the critical behavior of SGD will be in the mean-field universality class, given that the upper critical dimension for the Manna universality class is $d=4$ (a fact that was verified numerically for BRO~\cite{wilken2023dynamical}).We further conjecture that learning tasks in both artificial and biological neural networks involving manifold packing in high-dimensional representation spaces ($d \ge O(100)$) operate in an active phase where complete separation of manifolds without overlap is unachievable. Instead, partial overlap is necessary to fully utilize the representational capacity. This hypothesis is based on the exponential decay of the densest packing volume fraction, $\phi_v(d)$, for identical spheres with increasing dimension, $d$. The best-known upper bound of $\phi_v(d)$ in high dimensions, proved by Kabatyanskii and Levenshtein , scales as $\phi_v(d) \lesssim 2^{-0.599d}$ \cite{kabatiansky1978bounds,cohn2016conceptual}, showing the increasing sparsity of the densest packing.
Finally, our results offer a way of designing representation learning algorithms based on physical principles.

\begin{acknowledgments}
We particularly thank David J. Heeger for crucial comments on this work. We also thank Satyam Anand,  Shen Ai, Mathias Casiulis,  Paul M. Chaikin, Dov Levine, Flaviano Morone, Aaron Shih and Sam Wilken for valuable discussions. This work was supported by the National Science Foundation grant IIS-2226387, National Institute of Health under award number R01MH137669, Simons Center for Computational Physical Chemistry, and in part by the NYU IT High Performance Computing resources, services, and staff expertise.
\end{acknowledgments}
\bibliography{apssamp}

\appendix
\renewcommand{\theequation}{A\arabic{equation}}
\setcounter{equation}{0}
\onecolumngrid
\section{Appendix}
\twocolumngrid
\paragraph{\bf{Stochastic approximation of \particlewise BRO.}} 
In \particlewise BRO, the update rule for active particle $i$ at iteration $k$ reads,
\begin{equation}
    \begin{aligned}
        \mathbf{x}_{k+1}^{i} =\mathbf{x}_{k}^{i} + u^{i}_k \sum_{j \in \neighbour{i}} \Delta \mathbf{x}_k^{ji}.
     \label{eq:particle_bro_step}
    \end{aligned}
\end{equation}
where $\Delta \mathbf{x}_k^{ji} = -  \unitVec{j}{i}$. $u_k^{i}$ is a particle-wise noise sample drawn uniformly from $[0,\epsilon]$ at iteration k.

To derive a stochastic approximation, we follow the same procedure as for \pairwise BRO, decomposing the biased random kick into its mean value $\mathbf{g}_k^{i} = \mathbb{E}[u_k^i \sum_{j \in \neighbour{i}}\Delta \mathbf{x}_k^{ji} ] = \frac{1}{2} \sum_{j \in \neighbour{i}_k} \Delta \mathbf{x}_k^{ji}$ and the fluctuation around the mean $\mathbf{z}_k^{i} = u_k^i \sum_{j \in \neighbour{i}}\Delta \mathbf{x}_k^{ji} - \mathbf{g}_k^i$.  We then approximate the random fluctuation by the Gaussian random variable, $\boldsymbol{\eta}_i \sim \mathcal{N}(\mathbf{0}, \boldsymbol{\Sigma}^i)$ with covariance, $\mathbf{\Sigma}^i = \mathbb{E}[\mathbf{z}_k^i \mathbf{z}_k^{i^T}] = \frac{1}{3} \mathbf{g}_k^i {\mathbf{g}_k^i}^T$. Therefore, we approximate Eq.~\ref{eq:particle_bro_step} by
\begin{equation}
    \begin{aligned}
        \mathbf{x}_{k+1}^i =\mathbf{x}_{k}^i + \epsilon(\mathbf{g}_k^i + \boldsymbol{\eta}_k^i).
        \label{eq:particle_bro_approx}
    \end{aligned}
\end{equation}
By rewriting $\mathbf{g}_k^i = - \nabla^{i} V$, we obtain the stochastic approximation for the \particlewise BRO.
\begin{equation}
   \begin{aligned}
       \mathbf{x}_{k+1}^{i} &=  \mathbf{x}_{k}^{i} - \epsilon \nabla^{i} V + \frac{\epsilon}{\sqrt{3}} \sqrt{\boldsymbol{\Omega}^i} \cdot \boldsymbol{\xi}_k^i,
     \\ \text{where }
    \boldsymbol{\Omega}^i &= \nabla^{i} V \nabla^{i} V^T
   \end{aligned}
   \label{eq:particle-bro-sto-approx}
\end{equation}
The corresponding stochastic differential equation is, 
 \begin{equation}
      \mathrm{d}\mathbf{x}^{i}(t) = - \frac{\epsilon}{\tau} \nabla^{i} V \ \mathrm{d}t+ \frac{\epsilon}{\sqrt{3\tau}} \sum_{j} \sqrt{\boldsymbol{\Omega}^i(t)} \cdot \mathrm{d}\boldsymbol{W}^{ji}(t)
      \label{eq:particle-bro-sme}
 \end{equation}
 where $\mathbf{x}^{i}(t)$ and $\boldsymbol{\Omega}^{i}(t)$ are the continuous-time counterparts of the discrete-time variables in Eq.~\ref{eq:particle-bro-sto-approx} and $\tau$ is the time scale measuring the time elapsed between two discrete steps (see Supplemental Materials).

\paragraph{\bf Stochastic approximation of RCD}
The RCD update rule for active particle $i$ at timestep $k$ is
\begin{equation}
    \begin{aligned}
        \mathbf{x}_{k+1}^i = \mathbf{x}_k^i - \alpha \theta_k^i \nabla^i V,
    \end{aligned}
    \label{eq:particle-sgd}
\end{equation}
where 
\begin{equation*}
    \theta_k^i = 
    \begin{cases}
        1, \text{ if active particle $i$ is selected at iteration $k$} \\
        0, \text{ otherwise} 
    \end{cases}
\end{equation*}
Each active particles share the same probability, $b_f$, to be selected for the batch update, therefor $p(\theta_k^{i}=1)=b_f$ and  the mean value and the covariance of the gradient descent move are
\begin{equation}
   \begin{aligned}
       \mathbb{E}[-\alpha \theta_k^i \nabla^i V] &= - \alpha b_f \nabla^i V \\
       \mathrm{Cov}[-\alpha \theta_k^i \nabla^i V] 
       &= \alpha^2( b_f - b_f^2)\nabla^i V \nabla^i V^T
   \end{aligned}
\end{equation}
We approximate $-\alpha \theta_k^i \nabla^i V$ as
\begin{equation}
    -\alpha \theta_k^i \nabla^i V \approx -\alpha b_f \nabla^i V + \alpha \sqrt{b_f - b_f^2} \sqrt{\boldsymbol{\Omega}^i} \cdot \boldsymbol{\xi}^i_k
    \label{eq:particle-sgd-noise}
\end{equation}
such that the mean and covariance of $-\alpha \theta_k^i \nabla^i$V and its approximation (Eq.~\ref{eq:particle-sgd-noise}) are equal. 
Thus, the stochastic approximation of Eq.~\ref{eq:particle-sgd} is
\begin{equation}
     \mathbf{x}_{k+1}^i = \mathbf{x}_k^i -
     \alpha b_f \nabla^i V + \alpha \sqrt{b_f - b_f^2} \sqrt{\boldsymbol{\Omega}^i} \cdot \boldsymbol{\xi}^i_k
     \label{eq:rcd-sto-approx}
\end{equation}
and the corresponding stochastic differential equation is
 \begin{equation}
      \mathrm{d}\mathbf{x}^{i}(t) = - \frac{\alpha b_f}{\tau} \nabla^{i} V \ \mathrm{d}t+ \frac{\alpha}{\sqrt{\tau}}\sqrt{b_f - b_f^2} \sum_{j} \sqrt{\boldsymbol{\Omega}^i(t)} \cdot \mathrm{d}\boldsymbol{W}^{ji}(t).
      \label{eq:particle-sgd-sme}
 \end{equation}
 When $\alpha=\epsilon/b_f$ and $b_f=3/4$, the stochastic approximations of \particlewise BRO and SGD are equivalent.

\paragraph{\bf Stochastic approximation of SGD}
The SGD dynamics of one active pairs of particles $(i,j)$ at time $k$ is
\begin{equation}
    \begin{aligned}
    \mathbf{x}_{k+1}^{i} &= \mathbf{x}_{k}^{i} 
    - \alpha \theta^{ij}_k\nabla_i 
    U(\absDiff{i}{j})
    \\
    \mathbf{x}_{k+1}^{j}  &= \mathbf{x}_{k}^{j} -  \alpha \theta^{ij}_k \nabla_j 
    U(\absDiff{i}{j}) 
    \end{aligned}
    \label{eq:pairwise-sgd-dynamics}
\end{equation}
where $U(\absDiff{i}{j})>0$ only if $\absDiff{i}{j} < 2R$, with $R$ the particles radius, and 
\begin{equation*}
    \theta_k^{ij} = 
    \begin{cases}
        1, \text{ if active pair $(i,j)$ is selected at iteration $k$} \\
        0, \text{ otherwise} 
    \end{cases}
\end{equation*}
The active pairs share the same probability, $b_f$, to be selected for the batch update, therefore, $p(\theta_k^{ij}=1)=b_f$. Thus, the mean value and the covariance of the gradient descent move are
\begin{equation}
   \begin{aligned}
       \mathbb{E}[-\alpha \theta_k^{ij} \nabla^i U(\mathbf{x}^i_k - \mathbf{x}^j_k)] &= - \alpha b_f \nabla^i U(\mathbf{x}^i_k - \mathbf{x}^j_k) \\
       \mathrm{Cov}[-\alpha \theta_k^{ij} \nabla^i U(\mathbf{x}^i_k - \mathbf{x}^j_k)] &= \alpha^2(b_f - b_f^2)\nabla^i U^{ij} \nabla^i {U^{ij}}^T
   \end{aligned}
   \label{eq:pairwise-sgd-noise}
\end{equation}
In the same way as for RCD, we obtain the mean and covariance of $-\alpha \theta_k^{ij} \nabla^j U(\mathbf{x}^i_k - \mathbf{x}^j_k)$, yielding the stochastic approximation of a single pairwise interaction
\begin{equation}
   \begin{aligned}
       \mathbf{x}^i_{k+1} = \mathbf{x}^i_{k} - \alpha b_f \nabla^i U^{ij} + \alpha \sqrt{b_f - b_f^2} \sqrt{\boldsymbol{\Lambda}^{ji}} \cdot \boldsymbol{\xi}^{ji}_k \\
      \mathbf{x}^j_{k+1} = \mathbf{x}^j_{k}  - \alpha b_f \nabla^j U^{ij} + \alpha \sqrt{b_f - b_f^2} \sqrt{\boldsymbol{\Lambda}^{ij}} \cdot \boldsymbol{\xi}^{ij}_k
   \end{aligned}
    \label{eq:pairwise-sgd-sto}
\end{equation}
where $\boldsymbol{\Lambda}^{ji} = \nabla^i U^{ij} \nabla^i {U^{ij}}^T$ and $\boldsymbol{\Lambda}^{ij} = \nabla^j U^{ij} \nabla^j {U^{ij}}^T$. Since Eq.~\ref{eq:pairwise-sgd-dynamics} is reciprocal, the center of mass is conserved, $\mathbf{x}^i_{k+1} + \mathbf{x}^j_{k+1} = \mathbf{x}^i_{k} + \mathbf{x}^j_{k}$, with the reciprocity constraint, $\boldsymbol{\xi}^{ij}_k = -\boldsymbol{\xi}^{ji}_k$ on the noise. Then, adding the contribution of all the active pairs that overlap with particle $i$, we arrive at the stochastic approximation of SGD,
 \begin{equation}
    \mathbf{x}_{k+1}^i = \mathbf{x}_k^i - \alpha b_f \nabla^i V
     + \alpha \sqrt{b_f - b_f^2} \sum_{j \in \neighbour{i}_k} \sqrt{\boldsymbol{\Lambda}^{ji}} \cdot \boldsymbol{\xi}_k^{ji},
     \label{eq:pairwise-sgd-sto-approx} 
\end{equation} \raggedbottom

\paragraph{\bf Numerical comparison of \pairwise BRO, SGD, and their stochastic approximations.}

\begin{figure}[H]
\centering
\includegraphics[width=0.40\textwidth]{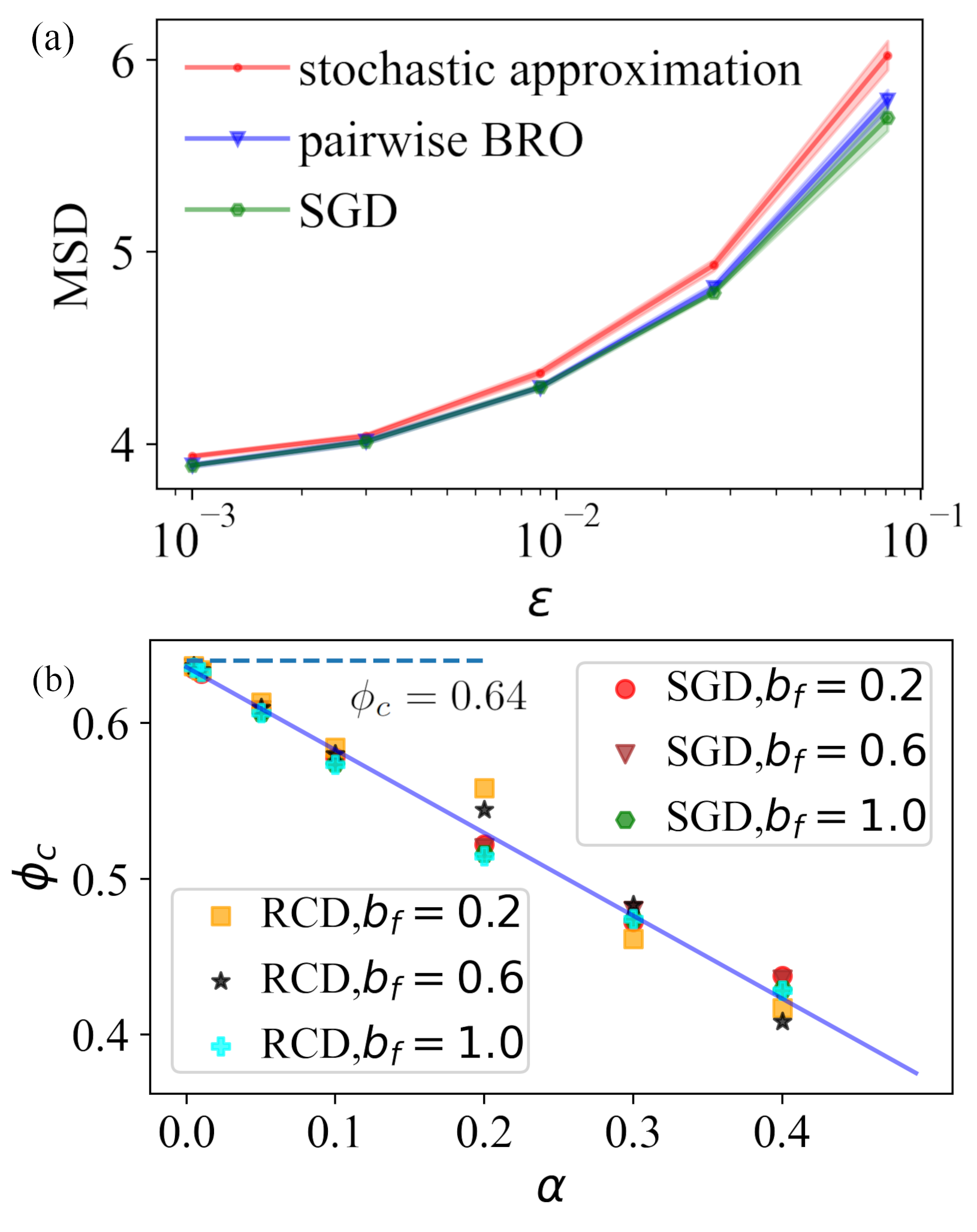}
    \caption{ (a) Mean square displacement (MSD) as a function of kick size ($\epsilon$) for SGD,pairwise BRO and their stochastic approximation measured at time step, $k=10^4$. For SGD and its stochastic approximation, we set $\epsilon = 4\alpha/3,~b_f = 3/4$ to match the BRO dynamics. (b) The critical packing fractions for SGD and RCD converge as $\phi_c \xrightarrow{} \phi_{RCP} \approx 0.64$ as $\alpha  \xrightarrow{} 0$. Both $\alpha$ and $\epsilon$ are measured in units of particle size, 2R.} 
    \label{fig:approx_phic}
\end{figure}

We evolve the stochastic approximation given by Eq.~\ref{eq:pairwise-sgd-sto-approx}, along with BRO and SGD, and compare their respective mean square displacements, $\text{MSD}(k)= \langle \frac{1}{N}\sum_i^N|\mathbf{x}^i_k - \mathbf{x}^i_0 |^2 \rangle$, at time step $k = 10^4$, for different schemes starting from the same initial configuration (Fig.~\ref{fig:approx_phic}(a)). We find that the MSD agrees between \pairwise BRO, SGD and their stochastic approximation for small kick sizes, while the stochastic approximation becomes less accurate for larger kick sizes, as expected (see Supplementary Material for the comparison between \particlewise BRO, RCD and their stochastic approximation). In addition, we find that the structure factor for BRO, SGD and RCD are consistent near the critical point (Fig.~{\SFigStructure} in Supplementary Material).
\paragraph{\bf Critical packing fraction for RCD and SGD}
In BRO, the critical packing fraction, $\phi_c$, which separates the absorbing from the active phase, approaches the random close packing fraction, $\phi_{RCP} \approx 0.64$ in three-dimension as the kick size $\epsilon \xrightarrow{} 0$. In Fig.~\ref{fig:approx_phic}(b) we show that all the dynamics converge to the same critical packing fraction, $\phi_c \xrightarrow{} \phi_{RCP}$ as $\epsilon \xrightarrow{} 0$ (or learning rate $\alpha  \xrightarrow{} 0$ for SGD/RCD). This result implies that $\phi_c(\epsilon \xrightarrow{} 0)$ is independent of both the noise level (quantified by $b_f$) and the reciprocity of the dynamics (whether \pairwise or \particlewise). This suggests that the specific noise characteristics of different schemes become less significant near the critical point.

\end{document}


\title{Absorbing state dynamics of stochastic gradient descent \\ (Supplementary Material)}
\author{Guanming Zhang}
\affiliation{\nyuphysics}
\affiliation{\nyusimons}
\email{gz2241@nyu.edu}
\author{Stefano Martiniani}
\affiliation{\nyuphysics}
\affiliation{\nyusimons}
\affiliation{\nyucourant}
\affiliation{\nyucns}
\email{sm7683@nyu.edu}
\newcommand{\nyuphysics}{Center for Soft Matter Research, Department of Physics, New York University, New York 10003, USA}
\newcommand{\nyuchemistry}{Department of Chemistry, New York University, New York 10003, USA}
\newcommand{\nyusimons}{Simons Center for Computational Physical Chemistry, Department of Chemistry, New York University, New York 10003, USA}
\newcommand{\nyucourant}{Courant Institute of Mathematical Sciences, New York University, New York 10003, USA}
\newcommand{\nyucns}{Center for Neural Science, New York University, New York 10003, USA}

\newcommand{\particlewise}{individual-particle }
\newcommand{\Particlewise}{Individual-particle }
\newcommand{\pairwise}{pairwise }
\newcommand{\Pairwise}{Pairwise }

\maketitle

\section{Algorithmic description for BRO, SGD and RCD}
\subsection{Algorithmic description of \pairwise BRO}
\begin{algorithm}[H]
\caption{\pairwise, reciprocal BRO}\label{alg:pair-bro}
\begin{algorithmic}
\State \textbf{Initialization:}
$\mathbf{x}_0^0,\mathbf{x}_0^1 ...\  \mathbf{x}_0^{N-1}$ are uniformly initialized in the periodic simulation box.
\\ set the time step, $T$ and kick size, $\epsilon$.
\For {k = 0 to T-1 }
    \For {each active pair, $(i,j)$ where $\absDiff{i}{j} < 2R$}
        \State $u_k^{ji} \gets \text{uniform}(0,\epsilon)$
        \State $u_k^{ij} \gets u_k^{ji}$
        \State $\mathbf{x}_{k+1}^i \gets \mathbf{x}_{k}^i - u_k^{ji} 
        \frac{\mathbf{x}_k^{j} - \mathbf{x}_k^i}{|\mathbf{x}_k^{j} - \mathbf{x}_k^{i} |}$
        \State $\mathbf{x}_{k+1}^j \gets \mathbf{x}_{k}^j - u_k^{ij} \frac{\mathbf{x}_k^{i} - \mathbf{x}_k^j}{|\mathbf{x}_k^{j} - \mathbf{x}_k^{i} |}$
    \EndFor
\EndFor
\end{algorithmic}
\end{algorithm}
\subsection{Algorithmic description of \particlewise BRO}
\begin{algorithm}[H]
\caption{\particlewise BRO}\label{alg:particle-bro}
\begin{algorithmic}
\State \textbf{Initialization:}
$\mathbf{x}_0^0,\mathbf{x}_0^1 ...\  \mathbf{x}_0^{N-1}$ are uniformly initialized in the periodic simulation box.
\\ set the time step, $T$ and kick size, $\epsilon$.
\For {k = 0 to T-1 }
    \For {each active particle $i$, where $|\neighbour{i}_k = \{j |\  |\mathbf{x}_k^j - \mathbf{x}_k^j| < 2R \}| > 0$  }
        \State $u_k^i \gets \text{uniform}(0,\epsilon)$
        \State $\mathbf{x}_{k+1}^i \gets \mathbf{x}_{k}^i - u_k^i \sum_{j \in 
 \neighbour{i}_k} \frac{\mathbf{x}_k^{j} - \mathbf{x_{k}}^i}{|\mathbf{x}_k^{j} - \mathbf{x}_k^{i} |}$ 
    \EndFor
\EndFor
\end{algorithmic}
\end{algorithm}

\subsection{Algorithmic description of energy-based SGD}

\begin{algorithm}[H]
\caption{energy-based stochastic gradient descent algorithm (SGD)}\label{alg:pair-sgd}
\begin{algorithmic}
\State \textbf{Initialization:}
$\mathbf{x}_0^0,\mathbf{x}_0^1 ...\  \mathbf{x}_0^{N-1}$ are uniformly initialized in the periodic simulation box
\\ set the time step, $T$, learning rate, $\alpha$, and batch size fraction, $b_f$.
\For {k = 0 to T-1}
    \For {each active pair, $(i,j)$ where $\absDiff{i}{j} < 2R$}
        \State $p \gets \text{uniform}(0,1)$
        \If {$p < b_f$}
            \State $\mathbf{x}_{k+1}^{i} \gets \mathbf{x}_{k}^{i} - \alpha \nabla_i 
    U(\absDiff{i}{j})$ 
            \State $\mathbf{x}_{k+1}^{j} \gets \mathbf{x}_{k}^{j} -  \alpha \nabla_j 
    U(\absDiff{i}{j})$ 
        \EndIf         
    \EndFor
\EndFor
\end{algorithmic}
\end{algorithm}

\subsection{Algorithmic description of RCD}
\begin{algorithm}[H]
\caption{random coordinate gradient descent algorithm (RCD)}\label{alg:particle-sgd}
\begin{algorithmic}
\State \textbf{Initialization:}
$\mathbf{x}_0^0,\mathbf{x}_0^1 ...\  \mathbf{x}_0^{N-1}$ are uniformly initialized in the periodic simulation box
\\ set the time step, $T$, learning rate, $\alpha$, and batch size fraction, $b_f$.
\For {$k = 0 \text{ to } T-1$}
    \For {each active particle $i$, where $|\neighbour{i}_k = \{j |\  |\mathbf{x}_k^j - \mathbf{x}_k^j| < 2R \}| > 0$  }
        \State $p \gets \text{uniform}(0,1)$
        \If {$p < b_f$} 
            \State $\mathbf{x}_{k+1}^i = \mathbf{x}_k^i - \alpha \nabla^i V$
        \EndIf
    \EndFor
\EndFor
\end{algorithmic}
\end{algorithm}

We use cell lists to implement all the algorithms above to accelerate finding the neighboring particles.

\section{ Continuous-time Stochastic differential equations}
To map the discrete-step stochastic dynamics to the continuous-time stochastic differential equation (SDE), 
we start from the following discrete-time dynamics,
\begin{equation}
    \mathbf{x}_{k+1} = \mathbf{x}_k + \tau \mathbf{a}  + \sqrt{\tau} \mathbf{B} \cdot \boldsymbol{\eta},
    \label{eq:discrete-dynamics}
\end{equation}
where the time scale, $\tau$, measures the continuous time elapsed as the discrete step jumps from $k$ to $k+1$ and $\boldsymbol{\eta} \sim \mathcal{N}(\mathbf{0},\mathbf{I})$ is a Gaussian random variable. We set the continuous time $t = k \tau$ and obtain,
\begin{equation}
    \mathbf{x}(k\tau + \tau) = \mathbf{x}(k\tau) + \tau \mathbf{a}  + \sqrt{\tau} \mathbf{B} \cdot \boldsymbol{\eta}
    \label{eq:EM-approx}
\end{equation}
where $\mathbf{x}(k\tau + \tau)$ and $\mathbf{x}(k\tau)$ are the continuous-time counterpart of $\mathbf{x}_{k+1}$ and $\mathbf{x}_k$.
It is important to note that the Euler-Maruyama update for a general continuous-time SDE, 
\begin{equation}
    \mathrm{d}\mathbf{x}(t) = \mathbf{a} \mathrm{d}t + \mathbf{B} \cdot \mathrm{d}\mathbf{W}(t),
    \label{eq:sde}
\end{equation}
is Eq.~\ref{eq:EM-approx}. Therefore, we can approximate discrete-time dynamics Eq.~\ref{eq:discrete-dynamics} by the continuous-time SDE, Eq.~\ref{eq:sde}.

\subsection{SDE for \pairwise BRO}
We rewrite the \pairwise SGD dynamics (Eq.~\EqParwiseBro) as
\begin{align}
    \mathbf{x}_{k+1}^{i} &=  \mathbf{x}_{k}^{i} -  \tau (\frac{\epsilon}{\tau}\nabla^{i}V) + \sqrt{\tau}(\frac{\epsilon}{\sqrt{3 \tau}} \sum_{j} \sqrt{\boldsymbol{\Lambda}^{ji}} )\cdot \boldsymbol{\xi}_k^{ji}    \label{eq:em-pairwise-bro}
\end{align}
Comparing this equation with Eq.~\ref{eq:EM-approx}, we get $\mathbf{a} = \frac{\epsilon}{\tau}\nabla^{i}V, ~\mathbf{B} = \frac{\epsilon}{\sqrt{3 \tau}} \sum_{j} \sqrt{\boldsymbol{\Lambda}^{ji}}$ and $\boldsymbol{\eta} = \boldsymbol{\xi}_k^{ji}$. Therefore, the SME for pairwise BRO is
\begin{equation}
     \mathrm{d}\mathbf{x}(t) = \frac{\epsilon}{\tau}\nabla^{i}V \mathrm{d}t +  \frac{\epsilon}{\sqrt{3 \tau}} \sum_{j} \sqrt{\boldsymbol{\Lambda}^{ji}} \cdot \mathrm{d}\mathbf{W}^{ji}(t)
\end{equation}
where the noise term satisfies the reciprocal relation, $\mathrm{d}\mathbf{W}^{ji}(t) = - \mathrm{d}\mathbf{W}^{ij}(t)$. This equation resembles the SDE for SGD updates \cite{li2019stochastic,stephan2017stochastic}.

\subsection{SDEs for SGD, \particlewise BRO and RCD }
By applying the same method, we attain the SDEs for \particlewise BRO,
\begin{equation}
    \mathrm{d}\mathbf{x}^{i}(t) = - \frac{\epsilon}{\tau}\nabla^{i} V \ \mathrm{d}t+ \frac{\epsilon}{\sqrt{3 \tau}} \sqrt{\boldsymbol{\Omega}^{i}(t)} \cdot \mathrm{d}\boldsymbol{W}^{i}(t),
\end{equation}
SGD,
 \begin{equation}
      \mathrm{d}\mathbf{x}^{i}(t) = - \frac{\alpha b_f}{\tau} \nabla^{i} V \ \mathrm{d}t+ \frac{\alpha}{\sqrt{\tau}}\sqrt{ (b_f - b_f^2)} \sum_{j} \sqrt{\boldsymbol{\Lambda}^{ji}(t)} \cdot \mathrm{d}\boldsymbol{W}^{ji}(t),
      \label{eq:pairwise-sgd-sme}
 \end{equation}
and RCD,
\begin{equation}
    \mathrm{d}\mathbf{x}^{i}(t) = -\frac{\alpha b_f}{\tau} \nabla^{i} V \ \mathrm{d}t+ 
    \frac{\epsilon}{\sqrt{\tau}} \sqrt{ (b_f - b_f^2)} \sqrt{\boldsymbol{\Omega}^{i}(t)} \cdot \mathrm{d}\boldsymbol{W}^{i}(t),
\end{equation}

\section{Energy fluctuations approximate the trace of the Hessian matrix}
For second order smooth potential $V$ we have that
\begin{align}
  \Delta V 
  &= \langle V(\mathbf{X}_{T} + \delta \mathbf{X}) - V(\mathbf{X}) \rangle_{\delta \mathbf{X}}\\
  &\approx  \langle\frac{\partial V}{\partial X_\alpha} \delta X_\alpha + \frac{1}{2} \frac{\partial^2 V}{\partial X_\alpha \partial X_\beta} \delta X_\alpha \delta X_\beta \rangle \\
  &=  \frac{\partial V}{\partial X_\alpha}  \langle \delta X_\alpha \rangle + \frac{1}{2} \frac{\partial^2 V}{\partial X_\alpha \partial X_\beta} \langle \delta X_\alpha \delta X_\beta \rangle
\end{align}
 As $\delta \mathbf{X} \sim \mathcal{N}(0,\sigma^2\mathbf{I})$, we have $\langle \delta X_\alpha \rangle = 0 $ and $\langle \delta X_\alpha \delta X_\beta \rangle = \sigma^2 \delta_{\alpha \beta}$. Therefore, we get the approximation,
 \begin{align}
     \Delta V =  \frac{\sigma^2}{2} \frac{\partial^2 V}{\partial X_\alpha \partial X_\alpha} \sim Tr(H)
 \end{align}
 where the Hessian matrix is $H_{\alpha \beta} = \frac{\partial^2 V}{\partial X_\alpha \partial X_\beta}$.

\section{Supplementary figures}

 \begin{figure}[h]
  \centering\includegraphics[width=0.98\textwidth]{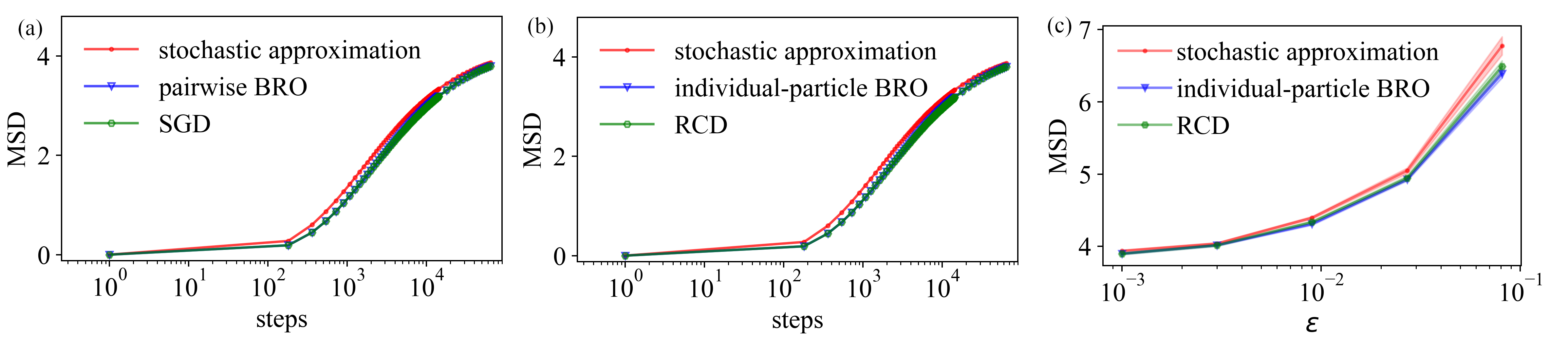}
    \caption{ MSD as a function of time steps at small kick size $\epsilon = 0.001 \times 2R $
    to show the equivalence among (a) SGD, \pairwise BRO and their stochastic approximation and (b)RCD, \particlewise BRO and their stochastic approximation. for (c)  RCD, \particlewise BRO and their stochastic approximation.
    MSD measured at fixed time step ($k=10000$) as a function of kick size. For SGD and RCD, the learning rate and batch size are set to match the BRO dynamics ($\alpha = \epsilon / b_f , b_f = 3/4$). Each curve is an average over 20 simulations with the same initial condition at $\phi=0.63$. $\epsilon$ is plotted in units of the particle diameter, $2R$. }
    \label{fig:msd_phi_0.63}
\end{figure}

\begin{figure}[H]
    \centering
    \includegraphics[width=0.55\textwidth]{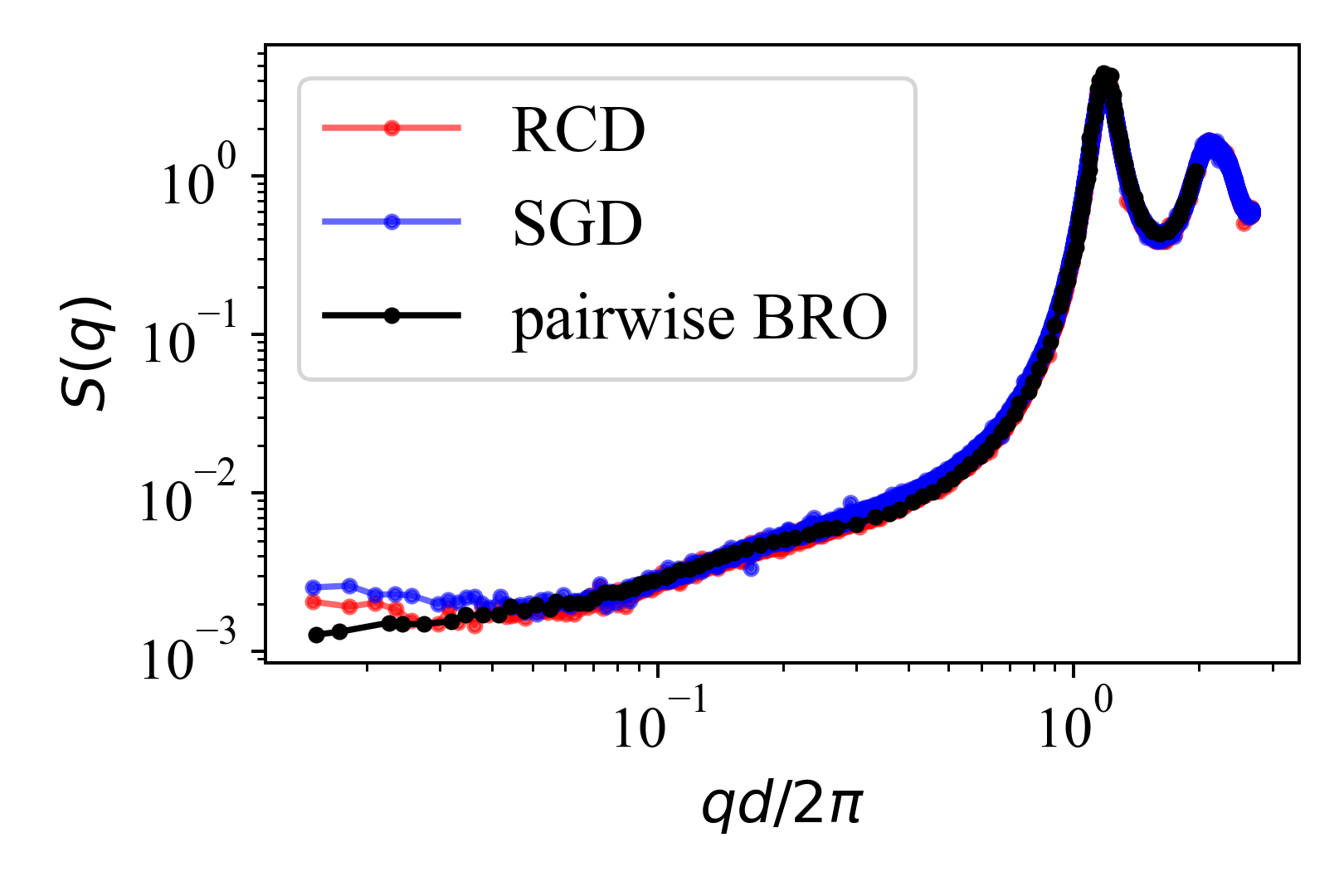}
    \caption{Structure factors for RCD, SGD and the BRO scheme described in \cite{wilken2021random} near the critical packing fraction. BRO data is from from Ref.~\cite{wilken2021random}.}
    \label{fig:enter-label}
\end{figure}

\begin{figure}[H]
    \centering\includegraphics[width=0.9\textwidth]{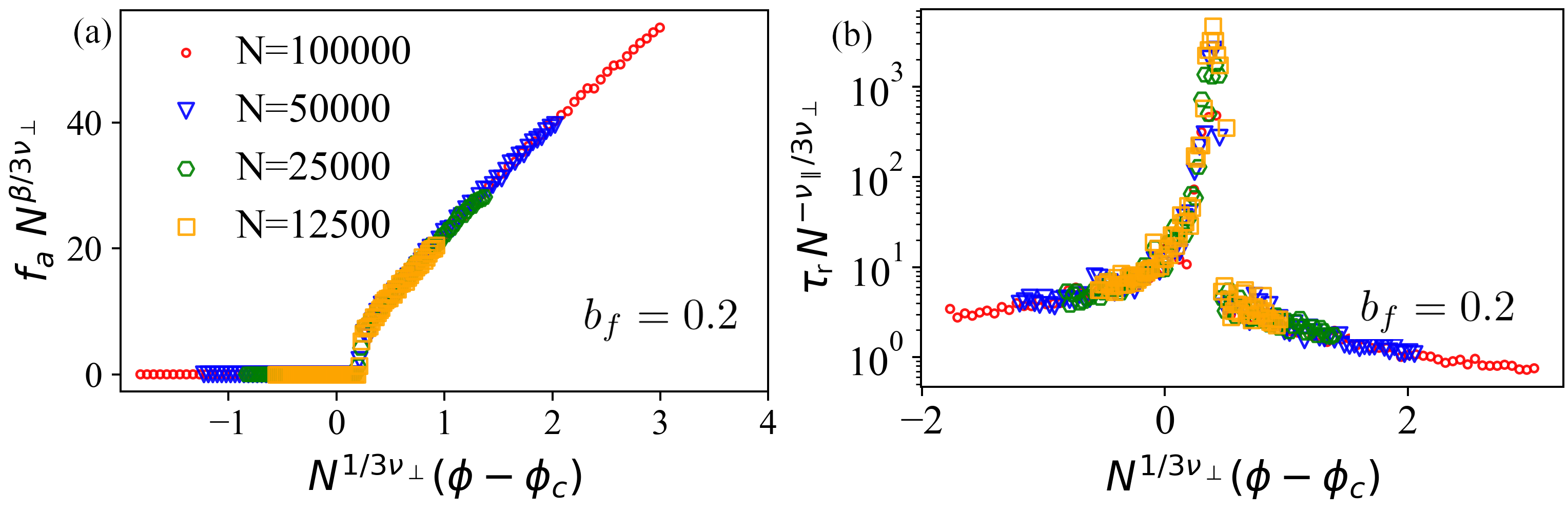}
    \caption{Finite size scaling analysis for RCD: (a) Steady-state activity as a function packing fraction for $b_f=0.2$. (b) Relaxation time as a function packing fraction for $b_f=0.2$ . $\nu_{\parallel} = 1.08, \beta = 0.84$ and $\nu_{\bot} = 0.59$ are Manna exponents in three-dimension.}
    \label{fig:exponents}
\end{figure}
\section{Simulation setup}
We perform simulations for $N=10^5$ particles (or $N=10^6$ particles for investigating the structure factors) in a cubic box of size $100 \times 100 \times 100$ with periodic boundary conditions. We use cell lists in the implementation of Algorithm~\ref{alg:pair-bro},~\ref{alg:particle-bro},~\ref{alg:pair-sgd} and \ref{alg:particle-sgd} for finding neighboring pairs. The particle radius are set to $ R= \sqrt[3]{\frac{3 \pi L^3 \phi}{4N}}$. Simulations are executed for $1 \times10^6$ to $8 \times 10^6$ steps or until an absorbing state is reached for $\phi < \phi_c$.

\bibliography{apssamp}